\documentclass[10pt, showpacs, prb, twocolumn, superscriptaddress]{revtex4}
\usepackage{graphicx}
\usepackage{amssymb}
\begin{document}
\title{Control of Correlations in Sr$_4$V$_2$O$_6$Fe$_2$As$_2$ by Chemical Stoichiometry}
\author{Athena S. Sefat}
\affiliation{Materials Science and Technology Division, Oak Ridge National Laboratory, Oak Ridge, TN, 37831, USA}
\author{David J. Singh}
\affiliation{Materials Science and Technology Division, Oak Ridge National Laboratory, Oak Ridge, TN, 37831, USA}
\author{V. Ovidiu Garlea}
\affiliation{Materials Science and Technology Division, Oak Ridge National Laboratory, Oak Ridge, TN, 37831, USA}
\author{Yuri L. Zuev}
\affiliation{Department of Physics, University of Tennessee, Knoxville, TN, 37996, USA}
\author{Michael A. McGuire}
\affiliation{Materials Science and Technology Division, Oak Ridge National Laboratory, Oak Ridge, TN, 37831, USA}
\author{Lindsay VanBebber}
\affiliation{Department of Materials Science and Engineering, University of Tennessee, Knoxville, TN, 37996}
\author{Brian C. Sales}
\affiliation{Materials Science and Technology Division, Oak Ridge National Laboratory, Oak Ridge, TN, 37831, USA}
\begin{abstract}
We show using a combination of powder X-ray and neutron diffraction, first principles
calculations, temperature- and field-dependent magnetization, heat capacity and resistivity data
that the superconducting behavior of `Sr$_4$V$_2$O$_6$Fe$_2$As$_2$' is dependent on synthesis conditions, particularly, heating profiles result in unintentional chemical doping. This compound can be tuned from a state in which the vanadium electrons are itinerant with a high electronic density of
states, to a state where the vanadium-oxide layers are insulating and presumably
antiferromagnetic. 
\end{abstract}
\maketitle

\section{Introduction}
Since the discovery of high-temperature superconductivity in iron-based materials,~\cite{Kamihara} five families of materials have been established. The ones about which most is known are the rare earth-based $R$FeAsO materials, the alkaline-earth-based $A$Fe$_2$As$_2$ compounds and the
monochalcogenides Fe(Se,Te). These are known as the 1111, 122, and 11 families, respectively.
The parents (i.e. the undoped stoichiometric compounds) are metallic, and often exhibit a spin-den\-sity-wa\-ve (SDW) type magnetic transition.~\cite{Rotter, Hsu} Superconductivity emerges when the SDW is suppressed, typically via chemical substitution at either the $R$, or $A$ or Fe-crystallographic site.~\cite{CWang, Ren1, Ren2, Sefat1, Sefat2, Sharma, Jiang, Yeh, Mizuguchi1, Mizuguchi2, Pitcher}

The two other families of iron-based superconductors are those of alkali-metal-based $A$FeAs and
oxygen-based Sr$_4T_2$O$_6$Fe$_2Pn_2$ ($T$ = transition metals; $Pn$ = pnictogens), the so-called 111 and 42622, respectively.~\cite{Pitcher, Ogino, Zhu} In 111 compounds, off-stoichiometry on the alkali-metal site or deliberate chemical doping yields superconductivity.~\cite{XCWang, Parker, Xia} In contrast to most of the other compounds, the undoped parents of 42622 family are reported to give superconductivity, with transition temperatures of $T_C = 17$ K for Sr$_4$Sc$_2$O$_6$Fe$_2$P$_2$~\cite{Ogino} and $T_C = 37$ K for Sr$_4$V$_2$O$_6$Fe$_2$As$_2$.~\cite{Zhu} Following those experimental reports, several theoretical papers have attempted to explain
the causes of such a high $T_C$ in Sr$_4$V$_2$O$_6$Fe$_2$As$_2$;~\cite{Shen, kwlee, mazin} one group reported not being able to reproduce superconductivity.~\cite{Pal} The present work is motivated by the experimental and theoretical controversies surrounding Sr$_4$V$_2$O$_6$Fe$_2$As$_2$. Specifically, why does a stoichiometric sample yield superconductivity? Is there a structural transition? Is there any type of magnetism or correlated electron behavior in the Sr$_4$V$_2$O$_6$ layer? Related to this we note that vanadium oxides with V$^{3+}$ in an octahedral environment show an extraordinary variety of behaviors related to the interplay of strong correlation effects, orbital orderings, magnetism, and itinerancy. In particular, one can tune through metal-insulators transitions. These are often accompanied by strong changes in magnetic order and lattice structure. Sr$_4$V$_2$O$_6$Fe$_2$As$_2$ combines perovskite-like, octahedrally-coordinated V$^{3+}$ oxide layers and FeAs-superconducting layers, and therefore potentially offers a window into the interplay between a correlated $3d$ oxide and a high-$T_C$ iron based
superconductor.

Like other fa\-mi\-lies of Fe-based su\-per\-con\-duc\-tors, Sr$_4$V$_2$O$_6$Fe$_2$As$_2$ has a quasi-two-di\-men\-si\-on\-al lay\-ered crystal structure, with iron in tetrahedral coordination with arsenic. The tetragonal symmetry of Sr$_2$FeO$_3$CuS-type ($P$4/nmm) combines FeAs layers with thick oxide layers giving the largest $d$-spacing ($\approx 15.7$ \AA) among the Fe-based superconductors. The report of $T_C = 37$ K gives the lattice parameters $a = 3.9296$ \AA\ and $c = 15.6732$ \AA,\cite{Zhu} while the report of non\-su\-per\-con\-duc\-ti\-vi\-ty
gives $a = 3.925(4)$ \AA\ and $c = 15.870(2)$ \AA.\cite{Pal} Based on this, one may infer
that the presence or absence of superconductivity depends on the slight changes in the structure,
due to unintended chemical doping. In iron-based superconductors, small structural effects have
been discussed in relation to $T_C$; they include the structures of a regular FeAs$_4$ tetrahedron
(109.47$^{\circ}$),~\cite{chlee} and the specific pnictogen height from the Fe-plane ($h_{\mathrm{Pn}} \approx 1.37$ \AA).~\cite{Mizuguchi3, Kuchinskii}

Here we report six sets of polycrystalline syntheses with different conditions, as well as physical
property results, in our attempts to produce `Sr$_4$V$_2$O$_6$Fe$_2$As$_2$.' Each synthesis method gives different physical properties behavior. We report an overview of syntheses followed by results
on the structure, thermodynamic and transport properties, and first-principles calculations. We
conclude that the parent (stoichiometric) 42622 may be magnetic and is not superconducting,
that unintentional chemical-doping (due to impurities) gives rise to superconductivity, that the $T_C$ value is connected with the $c$-lattice parameters which change with stoichiometry, and that there is no structural transition for the superconducting samples. We also note that the complex
behavior of 42622 is connected with the interplay between the V-O system and the FeAs layers.
Depending on composition, the material may cross from a non-superconducting, near-ferromagnetic
itinerant V$^{3+}$ state that is characterized by a much higher density-of-states, to a
superconducting state based on the insulating V$^{3+}$ layers.

\section{Experimental}

For the polycrystalline Sr$_4$V$_2$O$_6$Fe$_2$As$_2$ sample synthesis, ref.~\cite{Zhu} describes mixing stoichiometric amounts of SrAs, V$_2$O$_5$, SrO, Fe and Sr, pressing them into a pellet form, and then heating them in a sealed silica tube at $1150^{\circ}$C. Ref.~\cite{Pal} describes the use of a different set of reactants, namely V$_2$O$_5$, SrO$_2$, Sr, and FeAs, with heating at $750^{\circ}$C followed by heating at $1150^{\circ}$C. Here, we report a synthesis approach using another set of starting reactants, as we were unable to purify SrAs and avoided the use of highly air-sensitive Sr powder. Initially, SrCO$_3$ was decomposed to SrO by heating at $1325^{\circ}$C, and VAs was made through several heating steps in an evacuated silica tube (procedure: heat at $350^{\circ}$C, $600^{\circ}$C, and $800^{\circ}$C; regrind and heat at $800^{\circ}$C, and $1000^{\circ}$C; regrind and perform final anneal at $1000^{\circ}$C). Our starting reactants were SrO, VAs, Fe, and Fe$_2$O$_3$. They were mixed, pressed into pellets, and heated in evacuated silica tubes. Six pellets (labeled as Samples 1 through 6) were synthesized using two different heating profiles (a and b described below). The cooling of all reactions involved quenching by turning off the furnace.

We find that the superconducting behavior is dependent on the heating profile. A description of
each heating profiles follows. Profile (a) involved heating to an intermediate temperature in the
first synthesis step. The pellet was then reground, repressed into a pellet, and heated to a high
temperature in the second step. Using this outline, Samples 1, 2, and 3 were made. Sample 1 was
heated to $850^{\circ}$C (13 hrs), and then to $1150^{\circ}$C (24 hrs). Sample 2 was heated to $550^{\circ}$C (5 hrs), and then to $1150^{\circ}$C (40 hrs). Sample 3 was heated to $850^{\circ}$C  (13 hrs), and then to $1100^{\circ}$C (24 hrs). Profile (b) involved heating to a medium and a high temperature in the first synthesis step. The pellet was then reground, repressed, and heated to a high temperature in the next step(s). Samples 4, 5, and 6 were made using this profile. Sample 4 was heated to $850^{\circ}$C (12 hrs) and $1100^{\circ}$C (24 hrs), then to $1100^{\circ}$C (24 hrs), then to $1150^{\circ}$C (24 hrs). Sample 5 was heated to $850^{\circ}$C (15 hrs) and $1100^{\circ}$C (48 hrs), then to $1100^{\circ}$C (48 hrs). Sample 6 was heated to $850^{\circ}$C (15
hrs) and $1150^{\circ}$C (40 hrs), then to $1150^{\circ}$C (12 hrs).

The initial phase purity and structural identification were made via powder X-ray diffraction
using PANalytical X'pert PRO MPD. Lattice constants of all samples were determined at room
temperature from LeBail refinements using X'Pert HighScore Plus (2.2e version). For Sample 4,
powder x-ray diffraction data were collected at 11 K with Cu K$_{\alpha 1}$ radiation and an Oxford
Phenix closed-cycle cryostat.

Powder neutron diffraction measurements were performed on two samples (2 and 3), each
weighing $\sim 5$ g, using the HB2A diffractometer at the High Flux Isotope Reactor at Oak Ridge
National Laboratory.~\cite{Garlea} We used a monochromatic beam with a wavelength of $\lambda = 1.536$ \AA\ produced by a vertically focusing Ge(115) monochromator. A first set of measurements was
conducted on Sample 2, held in a cylindrical aluminum container and loaded in a $^4$He cryostat
(2-300 K). The aluminum holder was chosen to minimize the incoherent scattering and to
improve the detection of a potential magnetic scattering. Additional experiments were carried out
on Sample 3 that was loaded in a vanadium holder. Such measurements were focused on
investigating possible subtle changes of the lattice across the superconducting phase transition.
Rietveld refinements were performed for each set of data, at a given temperature, using the
FULLPROF program.~\cite{Rodriguez-Carvajal}

DC magnetization was measured as a function of temperature using a Quantum Design Magnetic
Property Measurement System (MPMS). For a temperature sweep experiment, the sample was
zero-field cooled (ZFC) to 1.8 K and data were collected by warming from 1.8 K in an applied
field. The sample was then field-cooled (FC) in the applied field, and the measurement was
repeated from 1.8 K. The magnetic-susceptibility results are presented per mole of formula unit
(cm$^3$/mol), $\chi_{\mathrm{m}}$.

The temperature and field dependence of DC electrical resistance were measured using a
Quantum Design Physical Property Measurement System (PPMS). Electrical contacts were
placed on samples in standard four-probe geometry by using Pt wires and Dupont silver paste. In
order to exclude the longitudinal contribution due to misalignment of Hall-voltage contacts, the
Hall resistivity was derived from the antisymmetric part of the transverse resistivity under
several magnetic field reversal (range of ± 6 Tesla) at a given temperature, i.e. $\rho_H = [\rho_H(+H)-\rho_H(-H)]/2$. The Hall coefficient ($R_H$) was then evaluated from $R_H =\rho_H/H$, and a linear fit through the different $H$ values. The temperature dependence of the heat capacity was also obtained using the PPMS, via the relaxation method.

First principles calculations were done using the general potential linearized augmented
planewave method~\cite{Singh} as implemented in the WIEN2k code~\cite{Blaha} and also an in-house code. We used the Perdew Burke Ernzerhof (PBE) generalized gradient approximation and the local spin
density approximation. We also did PBE+U calculations with the Coulomb parameter $U-J=6$ eV,
applied only to the vanadium. We based our calculations on the crystal structure of Zhu and coworkers.~\cite{Zhu}

\section{Results and Discussion}
\begin{figure}[t]
\centering
\includegraphics[width=0.75\columnwidth]{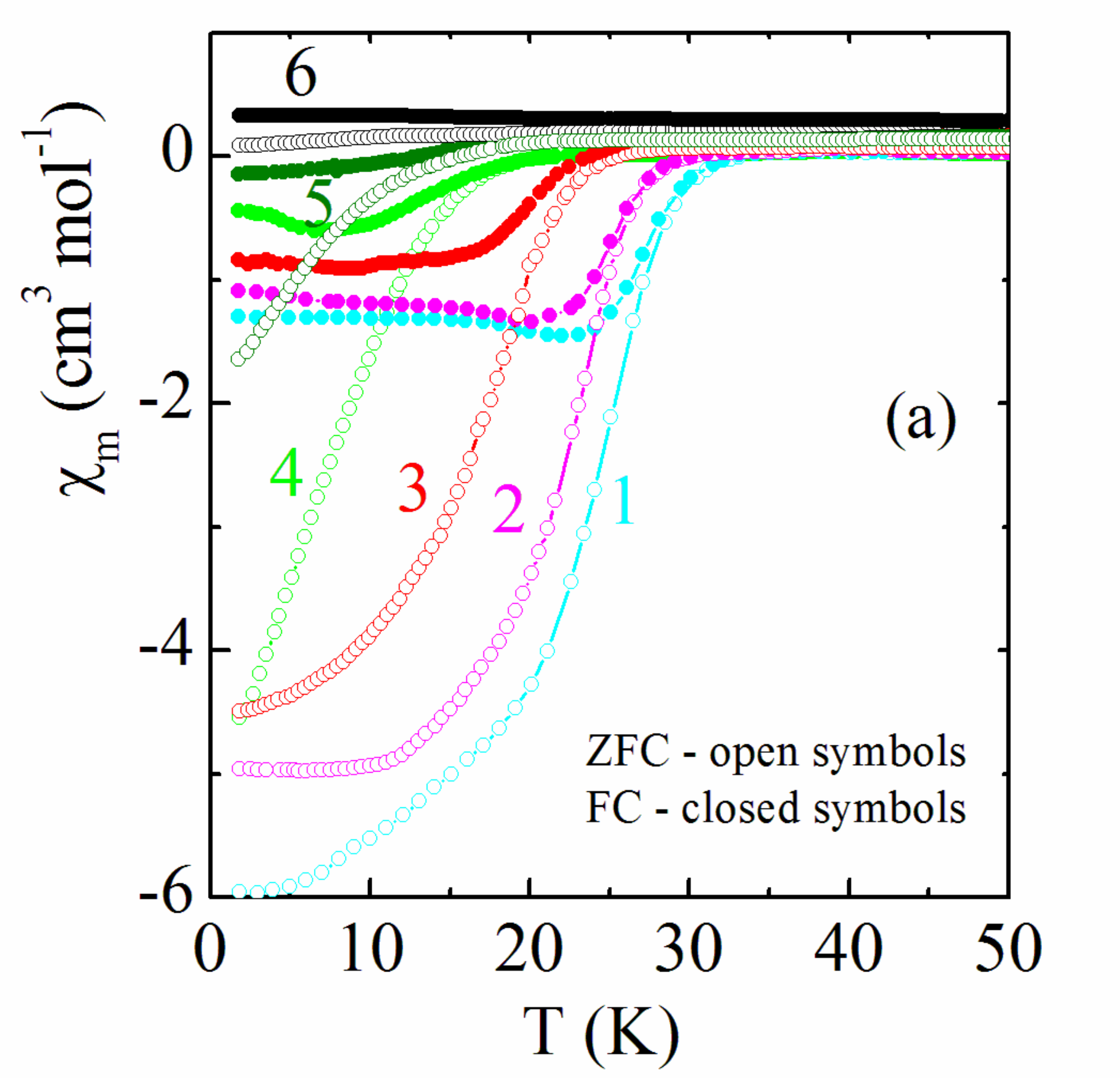}
\includegraphics[width=0.75\columnwidth]{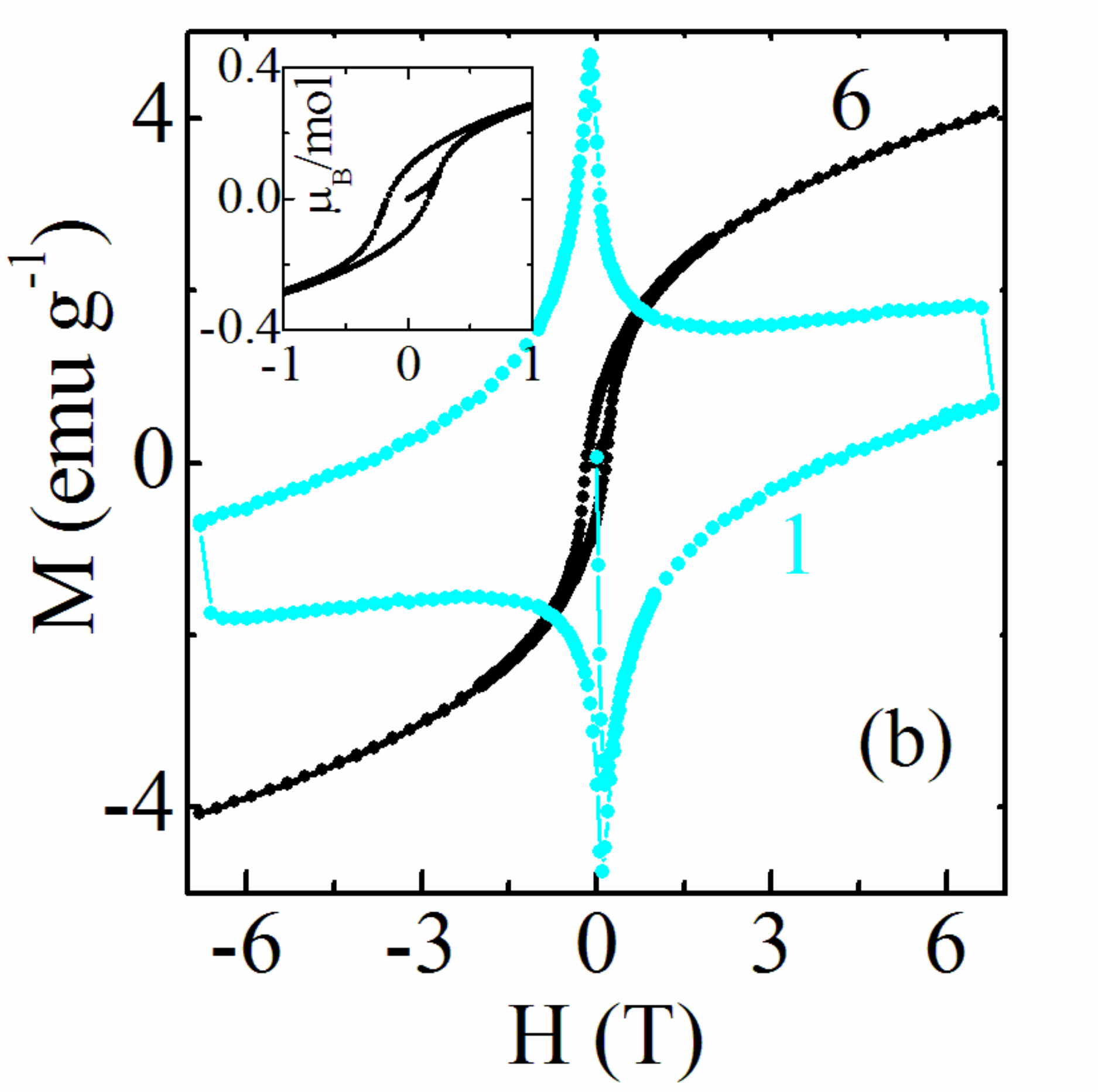}
\caption{(a) Temperature dependence of molar susceptibility $\chi_{\mathrm{m}}$ in zero-field-cooled (ZFC) and field-cooled (FC) modes for six differently prepared Sr$_4$V$_2$O$_6$Fe$_2$As$_2$ samples (20 or 50 G). (b) Field dependence of magnetization for Samples 1 and 6 at 2 K.}
\label{fig:suscept}
\end{figure}
Figure~\ref{fig:suscept}(a) shows the temperature dependence of magnetic susceptibility ($\chi_{\mathrm{m}}$) measured under ZFC and FC conditions at low magnetic fields (20 or 50 G) for Samples 1 through 6. $\chi_{\mathrm{m}}(T)$ becomes negative for Samples 1 to 5, indicating superconductivity. The ZFC and FC divergence temperatures ($\approx T_C$) for these samples are 33 K, 30 K, 25 K, 21 K, and 16 K, respectively. The highest $T_C$ value is consistent with the finding of the discovery paper for Sr$_4$V$_2$O$_6$Fe$_2$As$_2$, which reports $T_C = 31.5$ K from $\chi(T)$ results.~\cite{Zhu} Samples 1 and 2 have the highest $T_C$ values, as well as shielding and Meissner fractions. Assuming the theoretical density of  $\sim 5.6$ g/cm$^3$ and a $\chi$ value
of perfect diamagnetism, the shielding fractions are  $\sim 60$\% or 40\%, respectively, and the
Meissner fractions near the 10-15\% range at 2 K. In Samples 2 and 3, despite the high shielding
fraction of $\sim 40$\%, the Meissner fraction is only $\sim 4$\%. In Sample 4, both the shielding and Meissner fractions are small. However, the interpretation of Meissner fraction values, because
they are determined by pinning and penetration effects, may not be reliable on our
polycrystalline samples. In general, a large difference between Meissner and ZFC data indicates
strong pinning, and in such polycrystalline materials this may be due to grain boundaries or
impurities.
\begin{figure}[b]
\centering
\includegraphics[width=\columnwidth]{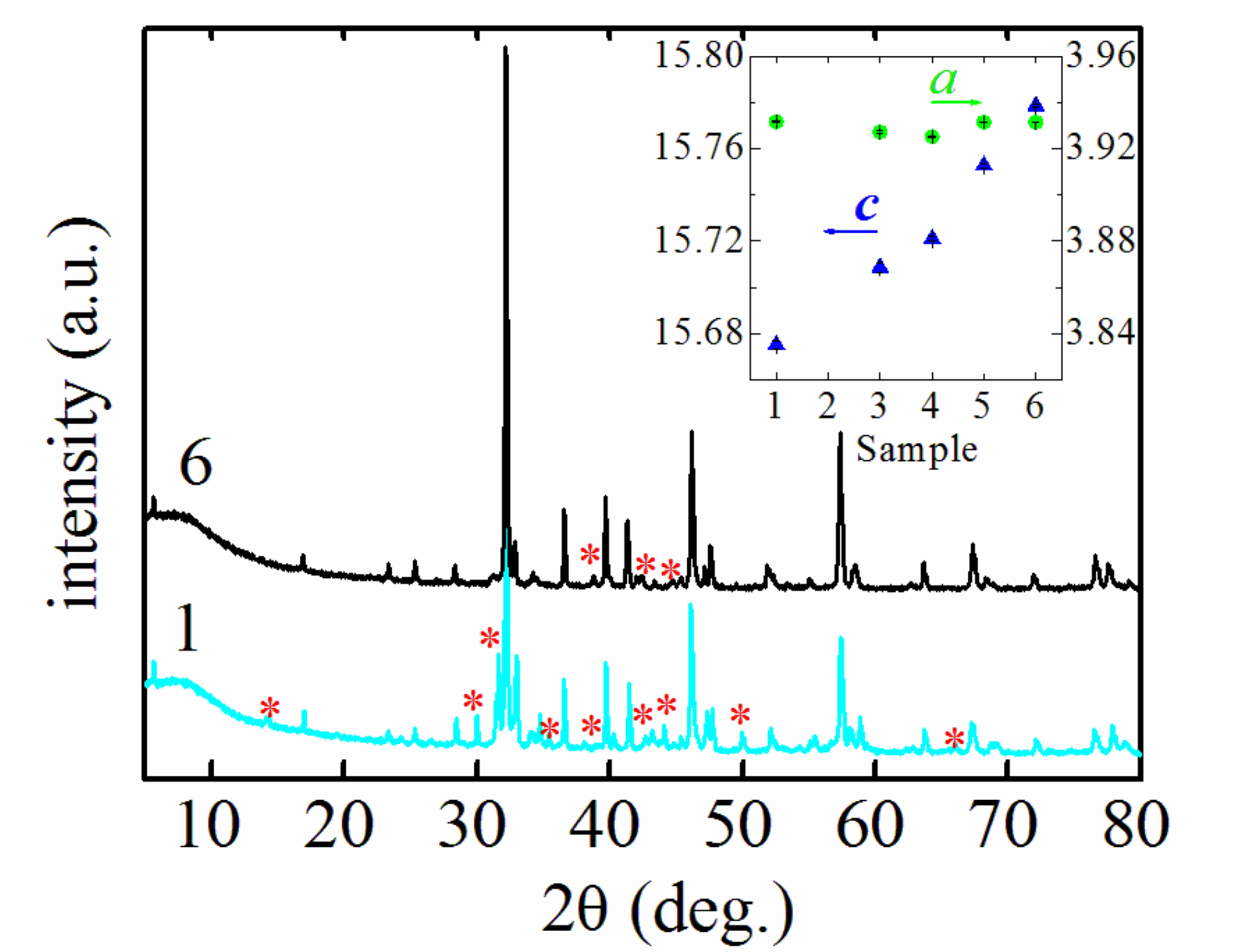}
\caption{X-ray diffraction data for Samples 1 and 6 at room temperature. Asterisks denote
impurity phases. Inset shows the change of lattice parameters with Sample.}
\label{fig:xray}
\end{figure}

For Samples 1 and 6, field-dependent magnetization data at 2 K are shown in Figure~\ref{fig:suscept}(b). For Sample 1, the hysteresis width and the sample size of about 1 mm translates to $J_C$ of the order of few kA/cm$^2$. This low value suggests that the grains are not coupled, and the persistent current circulates around each grain individually, rather than the sample as a whole. Sample 6, with a positive value of $\chi_{\mathrm{m}}$ and weak divergence of ZFC-FC at $\sim 15$ K (Fig.~\ref{fig:suscept}a), shows a weak ferromagnetic signal at 2 K (Fig. 1b), which exceeds the small superconductive diamagnetic signal. The magnetic hysteresis for Sample 6 has a remanent magnetization value of $M_r = 0.1\mu_B$/mol of f.u. and a coercive field of 0.2 Tesla (inset of Fig. 1b).

The room-temperature X-ray powder-diffraction data identified the main phase of all samples as
Sr$_2$FeO$_3$CuS-structure-type ($P4/nmm$). The diffraction data for Samples 1 and 6 are presented in
Figure~\ref{fig:xray}. Sample 1, a bulk superconductor with the highest $T_C$, has the greatest number of impurity peaks (shown by asterisks). In comparison, Sample 6, a non-bulk
superconductor has a cleaner diffraction pattern. The impurity phases are expected to produce
elemental deficiencies in the main phase that would force the stoichiometry away from 42622 in
Sr$_4$V$_2$O$_6$Fe$_2$As$_2$. Using a Le Bail fit, the lattice constants of Sample 1 are refined as $a = 3.9316(3)$ \AA~and $c = 15.675(2)$\AA, and for Sample 6 as $a = 3.9314(1)$ \AA\ and $c = 15.779(1)$ \AA. In fact, the $c$-lattice parameter increases with decreasing $T_C$ (sample number), while the $a$-lattice parameter does not show much change. The change of lattice parameters for the series of Samples is shown in inset of Figure~\ref{fig:xray}. Superconductivity may be sensitive to small structural changes and correlated with smaller $c$-lattice parameter. Our results are consistent with previous reports on Sr$_4$V$_2$O$_6$Fe$_2$As$_2$ that yields a $T_C = 30$ K for a sample with $c = 15.6732$ \AA,\cite{Zhu} and no superconductivity for a sample with $c = 15.870(2)$ \AA.\cite{Pal} Reports find the arsenic height ($h_{\mathrm{As}}$) from the Fe-layer of $\approx 1.38$ \AA~is crucial for achieving maximum $T_C$.\cite{Mizuguchi3, Kuchinskii} For Sr$_4$V$_2$O$_6$Fe$_2$As$_2$, $h_{\mathrm{As}}$ is reported as 1.42 \AA~and pressure experiments have reported a higher $T_C$ of 46 K under 4 GPa;\cite{Kotegawa} these confirm the importance of smaller $c$-lattice parameter.
\begin{figure}[t]
\centering
\includegraphics[width=\columnwidth]{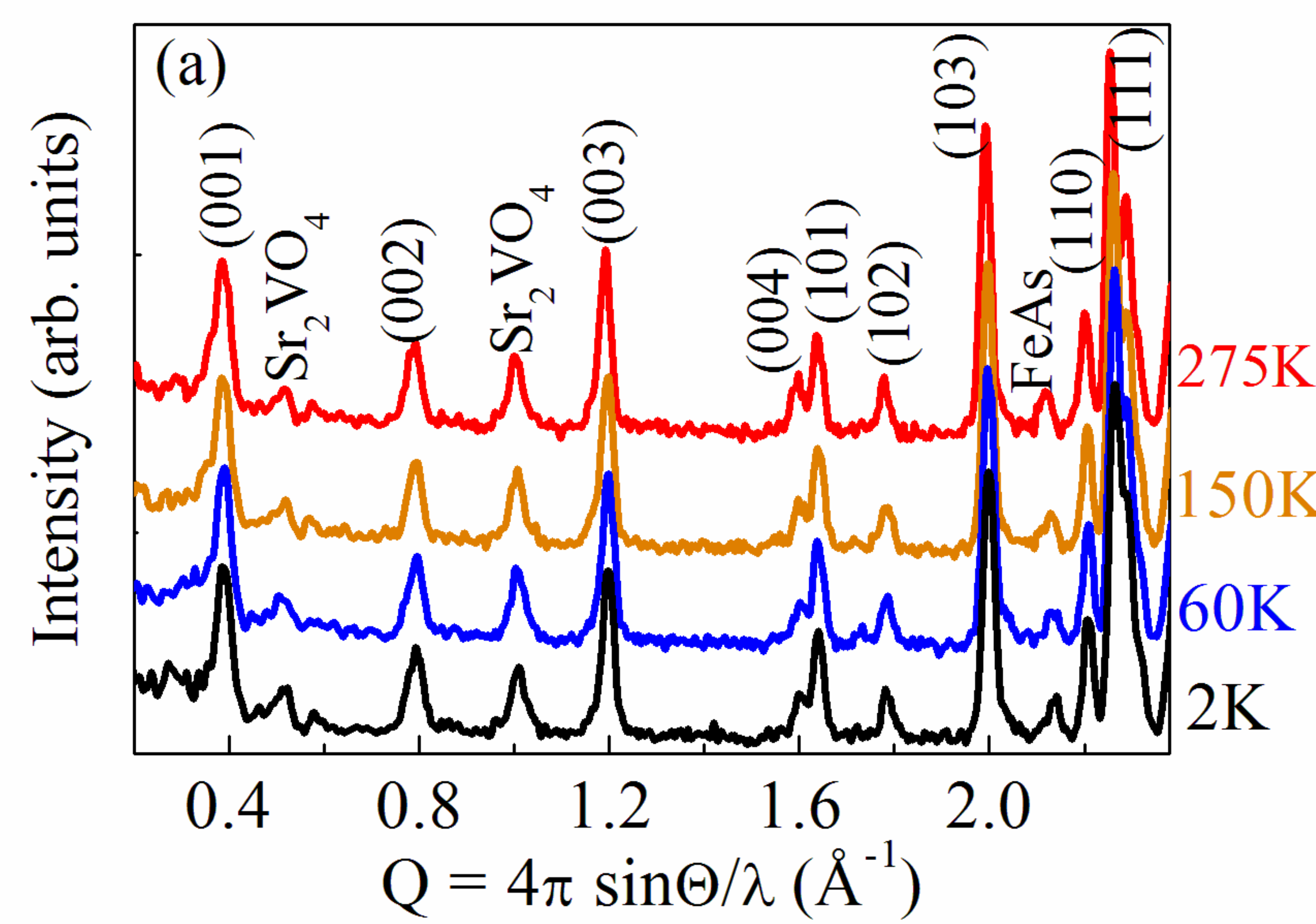}\\
\includegraphics[width=\columnwidth]{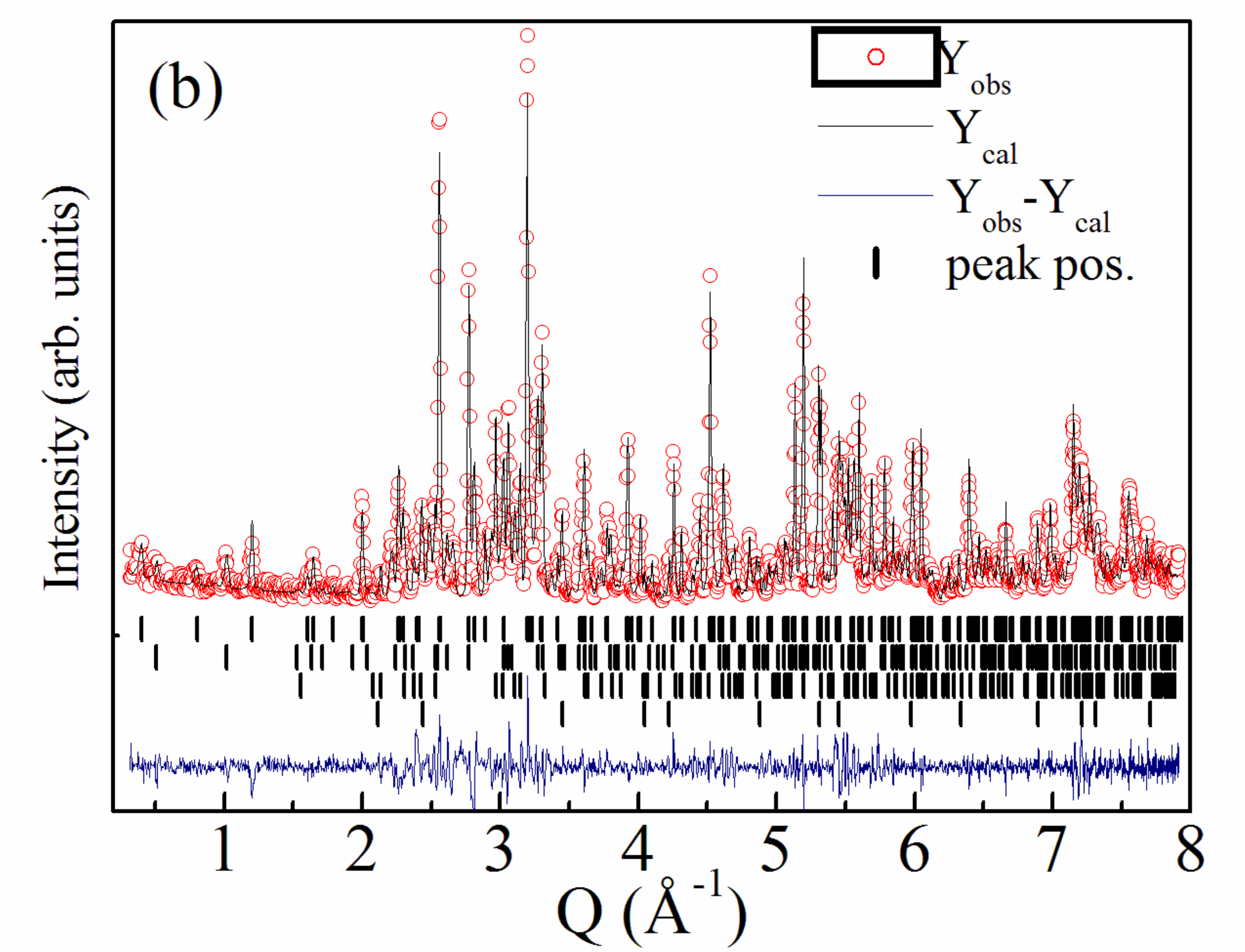}
\caption{The neutron powder diffraction data collected on the superconducting samples. (a)
Measurements at various temperatures on Sample 2, indicating no significant change of the
scattering; (b) Typical Rietveld refinement plot for Sample 3, with vertical bars denoting the peak
position of the main phase (42622) and impurity phases (Sr$_2$VO$_4$, FeAs, SrO).}
\label{fig:neut}
\end{figure}
In order to examine eventual phase transitions in the superconducting samples, neutron powder
diffraction measurements were performed on Samples 2 and 3. A comparison of the data at four
temperatures (2 K, 60 K, 150 K and 275 K) is shown in Figure~\ref{fig:neut}(a). The data are shifted vertically so they can be better visualized. No structural transformations were detected or magnetic long range order within the instrument detection limit ($0.3\mu_B$/f.u. for an antiferromagnet). Also, powder x-ray diffraction data (not shown) confirm the absence of structural transitions for Sample 4 down to 11 K, giving $c/a$ = 3.963 at 11 K and $c/a$ = 3.986 at 300 K. These values are comparable to neutron diffraction data (Fig.~\ref{fig:ac}).

\begin{figure}[t]
\centering
\includegraphics[width=0.7\columnwidth]{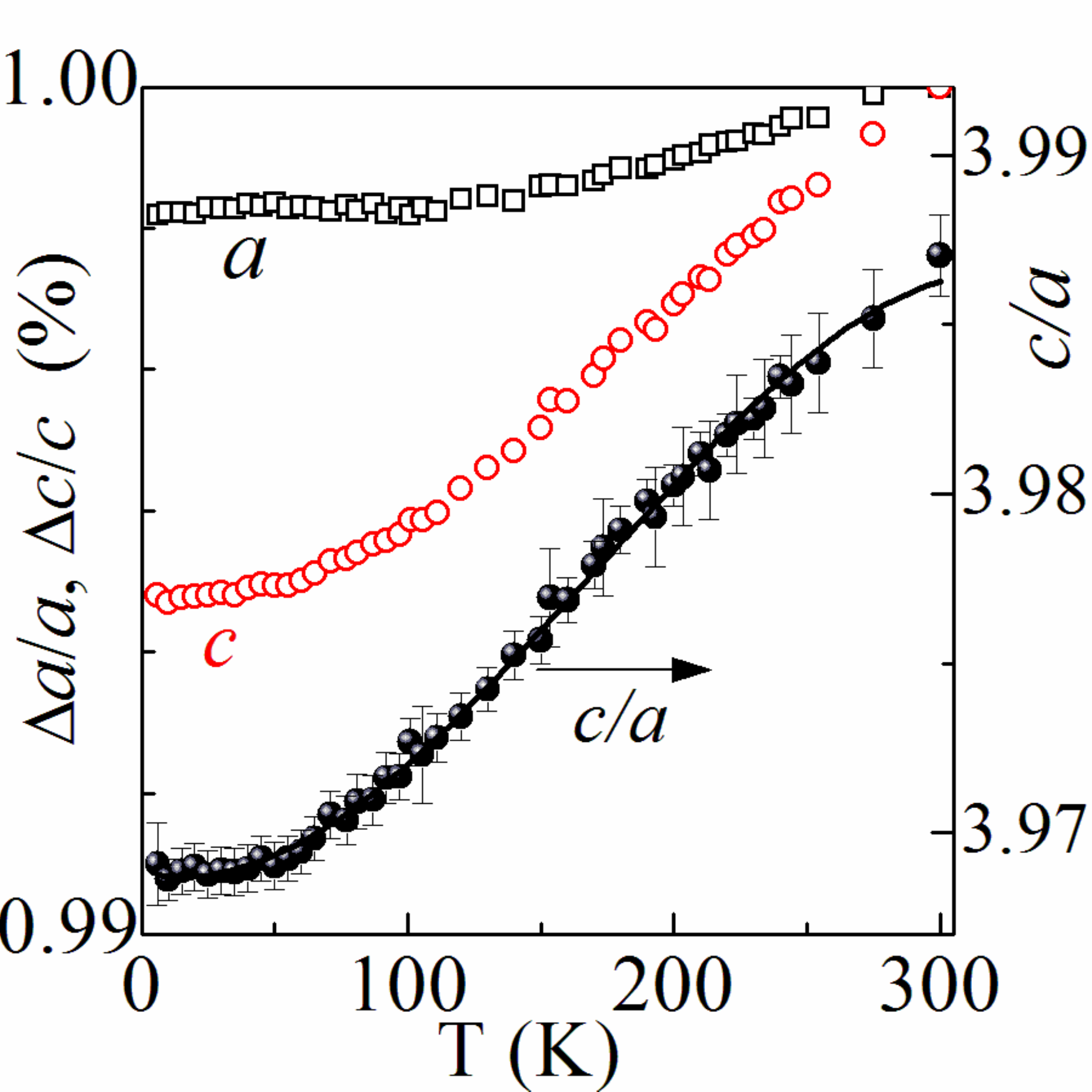}
\caption{Temperature dependence of lattice parameters, $a$ and $c$, as determined from the refinement
of neutron powder diffraction data for Sample 3. The ratio $c/a$ follows a monotonic behavior and is
well described by a cubic polynomial function.}
\label{fig:ac}
\end{figure}
\begin{figure}[t]
\centering
\includegraphics[width=0.75\columnwidth]{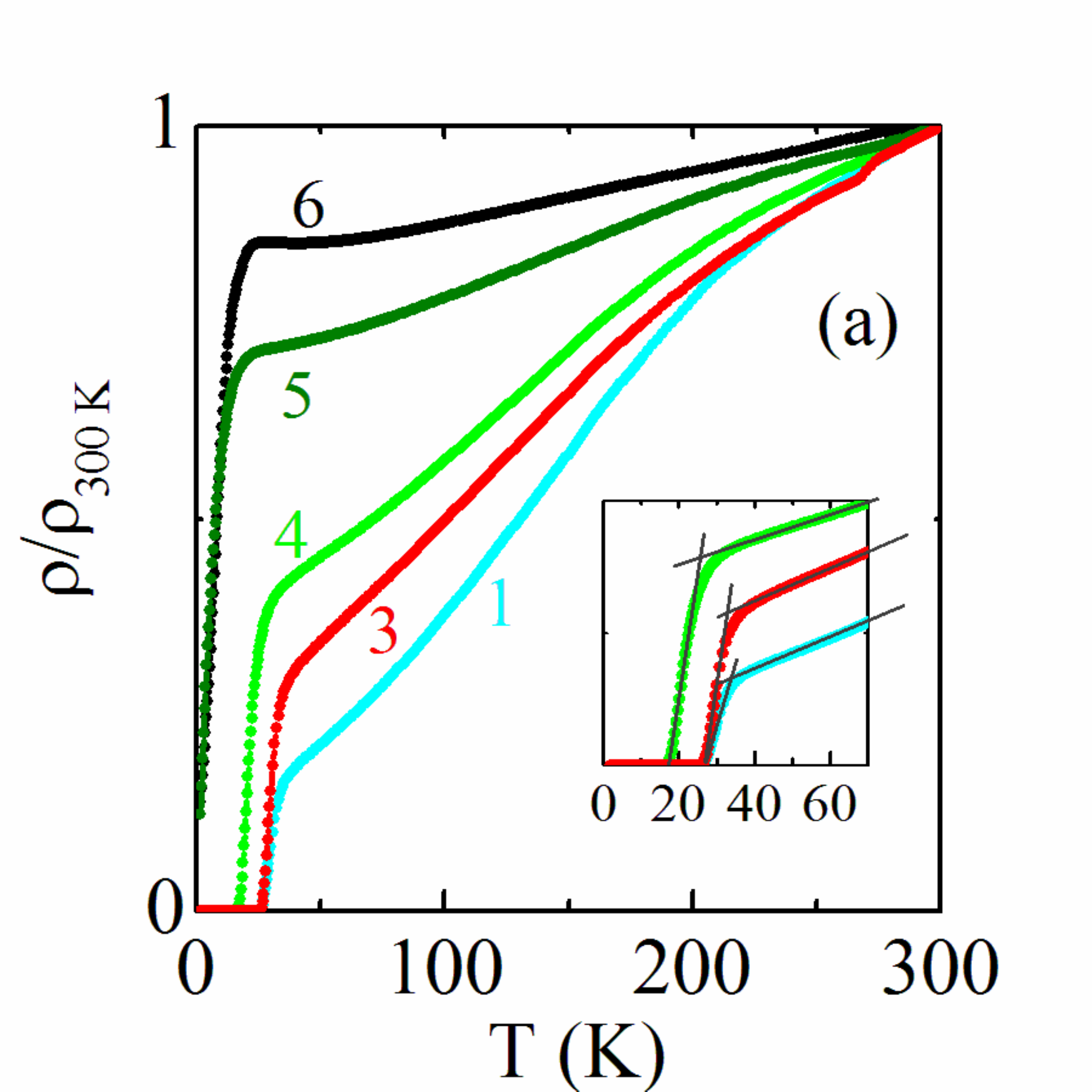}
\includegraphics[width=0.75\columnwidth]{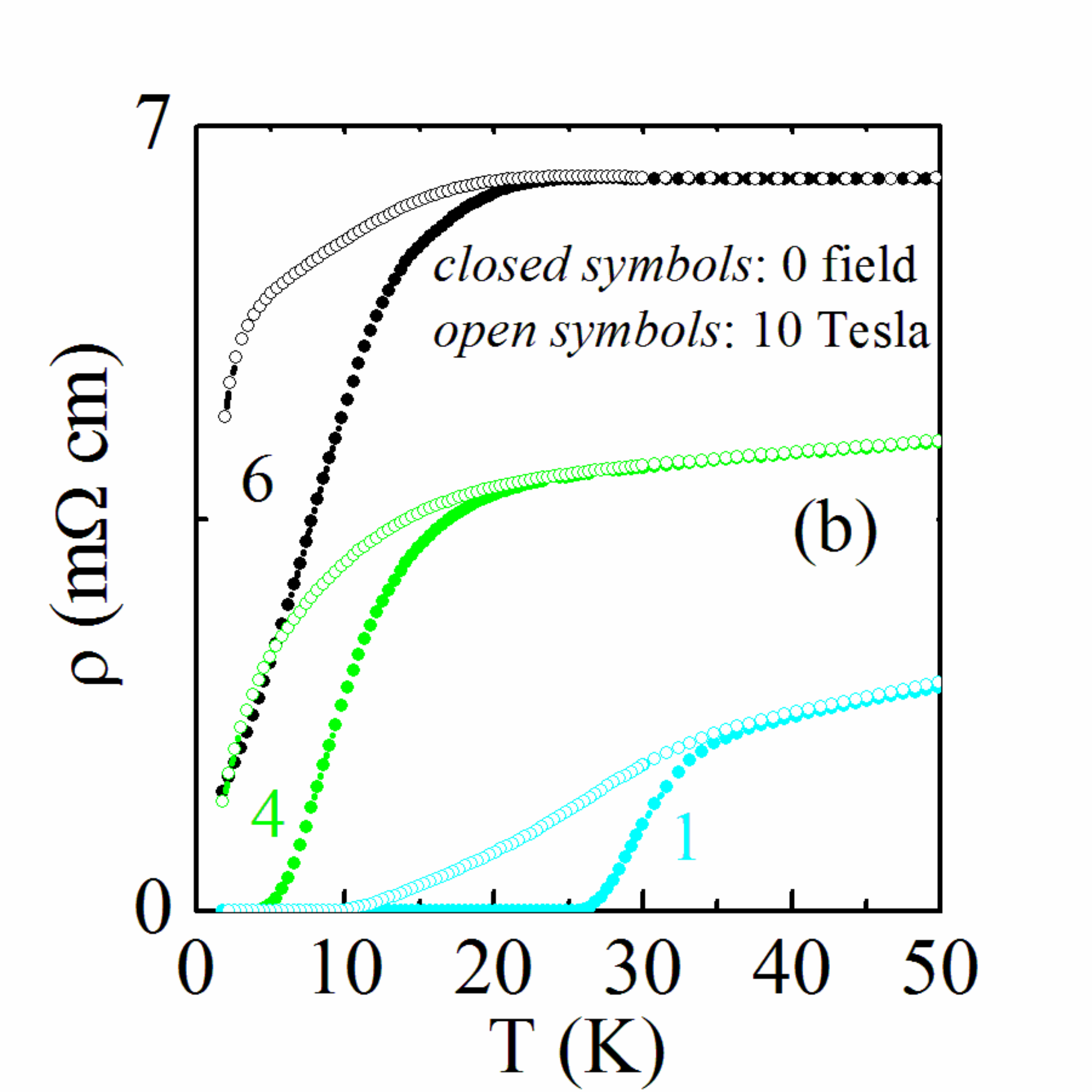}
\caption{(a) Temperature dependence of normalized electrical resistivity measured upon cooling
from room temperature to 1.8 K, at zero field. Inset is the enlarged part of low-temperature data illustrating the shift of the onset of superconducting transition temperature for 4. (b) Temperature dependence of resistivity below 50 K, at zero and 10 T applied magnetic fields.}
\label{fig:R}
\end{figure}

Additional neutron powder diffraction experiments were carried out on Sample 3, for which
measurements were focused on investigating possible subtle changes of the lattice across $T_C$.
Rietveld refinements were performed for each set of data (at a given temperature) using the
FULLPROF program. We were able to detect a number of impurity phases corresponding to
Sr$_2$VO$_4$ ($\sim 6$\%), FeAs ($\sim 9$\%), SrO ($\sim 4$\%), as well as a few unidentified Bragg reflections. The latter have been found to be well described by a tetragonal unit cell ($a = 7.5861$\AA, $c = 6.6952$ \AA). For the structural refinement of the main phase, the site occupancies for all atoms were maintained fixed to agree with the nominal chemical composition and the isotropic thermal parameters were constrained to be the same for all similar atoms. A typical Rietveld refinement plot from neutron diffraction data is shown in Figure~\ref{fig:neut}(b). Similar to Sample 2, no additional reflections were observed across the range of measured temperatures (4 K to room temperature).
In a recent x-ray and neutron study by Tegel and coworkers~\cite{Tegel} it has been suggested that 7\% vanadium-doping is responsible for the suppression of superconductivity in 42622. We believe
that such conclusions cannot be substantiated by neutron data since the scattering from vanadium
is almost entirely incoherent. Also in this report,~\cite{Tegel} the $c$-lattice parameter of the super- and non-superconductive phases was found to remain unchanged, which is in disagreement to earlier reports~\cite{Zhu, Pal} and our findings, although this may relate to different synthesis conditions.

Figure~\ref{fig:ac} displays the temperature dependence of the relative lattice parameters ($\Delta a/a$ and $\Delta c/c$) of Sample 3. One can observe a quite significant difference in the thermal behavior of the lattice spacing along the $a$ and $c$ crystallographic directions. The $c$ parameter follows a monotonous trend with decreasing temperature while the $a$ constant decreases less dramatically and flattens near 100 K. The $c/a$ ratio was further used to quantify the evolution of the lattice distortion vs. temperature. This ratio is obviously dominated by and follows a similar trend to the $c$-lattice parameter. The trend is well described by a cubic polynomial function, indicated by solid line in Figure~\ref{fig:ac}

For all samples, the room temperature electrical resistivity $\rho_{\mathrm{300\:K}}$ ranged from 3 to 8 m$\Omega\cdot$cm. Figure~\ref{fig:R}(a) shows the temperature-dependence of resistivity down to 1.8 K, normalized to room temperature. Samples 1, 2 and 4 show a plateau-like shape, in the 150 K to 200 K region, while Samples 5 or 6 do not. This feature was observed in the first report of the superconducting Sr$_4$V$_2$O$_6$Fe$_2$As$_2$ and was suggested to be the result of an incomplete suppression of an antiferromagnetic order.~\cite{Zhu} Our neutron diffraction data here shows no signs of magnetic order for Sample 2 (Fig.~\ref{fig:neut}), however, features are noted in heat capacity (Fig.~\ref{fig:Cp}).

Samples 1, 2 and 4 have zero resistance at low temperatures, while Samples 5 and 6 do not. This suggests that in the latter two samples, individual grains become superconducting on cooling, but do not form a continuous path for current conduction. For Samples 1, 2 and 4, the onset transition temperatures (illustrated in inset of Fig. 5a) for 90\% of the normal-state value are $T_C^{\mathrm{
onset}}\approx 34$ K, 32 K and 25 K, respectively. The lowest measured resistivity for Samples 5 and 6 are $\rho_{\mathrm{1.8\: K}} = 0.8$ m$\Omega$cm and 1.1 m$\Omega$cm, respectively. The transition widths for the superconducting samples are broad, $\Delta T_C \approx 7$ K, indicative of poor sample homogeneity. This value is larger than other Fe-based superconductors such as BaFe$_{1.8}$Co$_{0.2}$As$_2$ (0.6 K),\cite{Sefat2} LaFe$_{0.92}$Co$_{0.08}$AsO ($\Delta T_C \approx 2$ K),\cite{Sefat1} and LaFeAsO$_{0.89}$F$_{0.11}$ ($\Delta T_C \approx 4.5$ K).\cite{Sefat3}

The effects of 10 Tesla applied field on $\rho(T)$ are shown in Figure~\ref{fig:R}(b). For Sample 1, the $T_C$ onset is only slightly affected and the resistance moves eventually to zero after that. The $R=0$ temperature is suppressed by $\sim 15$ K in sample 1, showing large vortex liquid regime together with the effect of percolation, whereby the magnetic field quenches some of the grains that were available for conduction in zero-field. This effect is pronounced most for such anisotropic materials.  From the shift of $T_C^{\mathrm{onset}}$ we can determine the slope $dH_{C2}/dT$ near $T_C$, although the result depends on
how we quantify the $T_C^{\mathrm{onset}}$. Using the two tangents method, we obtain $dH_{C2}/dT\sim -9$ T/K using four field values, in close agreement with ref.~\cite{Zhu}, and the use of Werthamer-Helfand-Hohenberg (WHH) equation gives $H_{C2}(0) = 0.69T_C|dH_{C2}/dT| \approx 200$ T. This large $H_{C2}(0)$ value together with the largest gap
parameter $2\Delta/k_BT_C \approx 7$ reported in the Fe-based superconductor~\cite{Hiraishi} could be taken as an evidence that the paramagnetic (Chandrasekhar-Clogston) limit is exceeded in 42622. On the other hand, if we define $T_C^{\mathrm{onset}}$ to present a fixed 90\% of the normal-state value ($\rho_N$) at each field $H$, we obtain a smaller value of $dH_{C2}/dT \approx -3$ T/K and $H_{C2}(0) \approx 70$ T, which does not exceed the paramagnetic limit.

\begin{figure}[b]
\centering
\includegraphics[width=0.75\columnwidth]{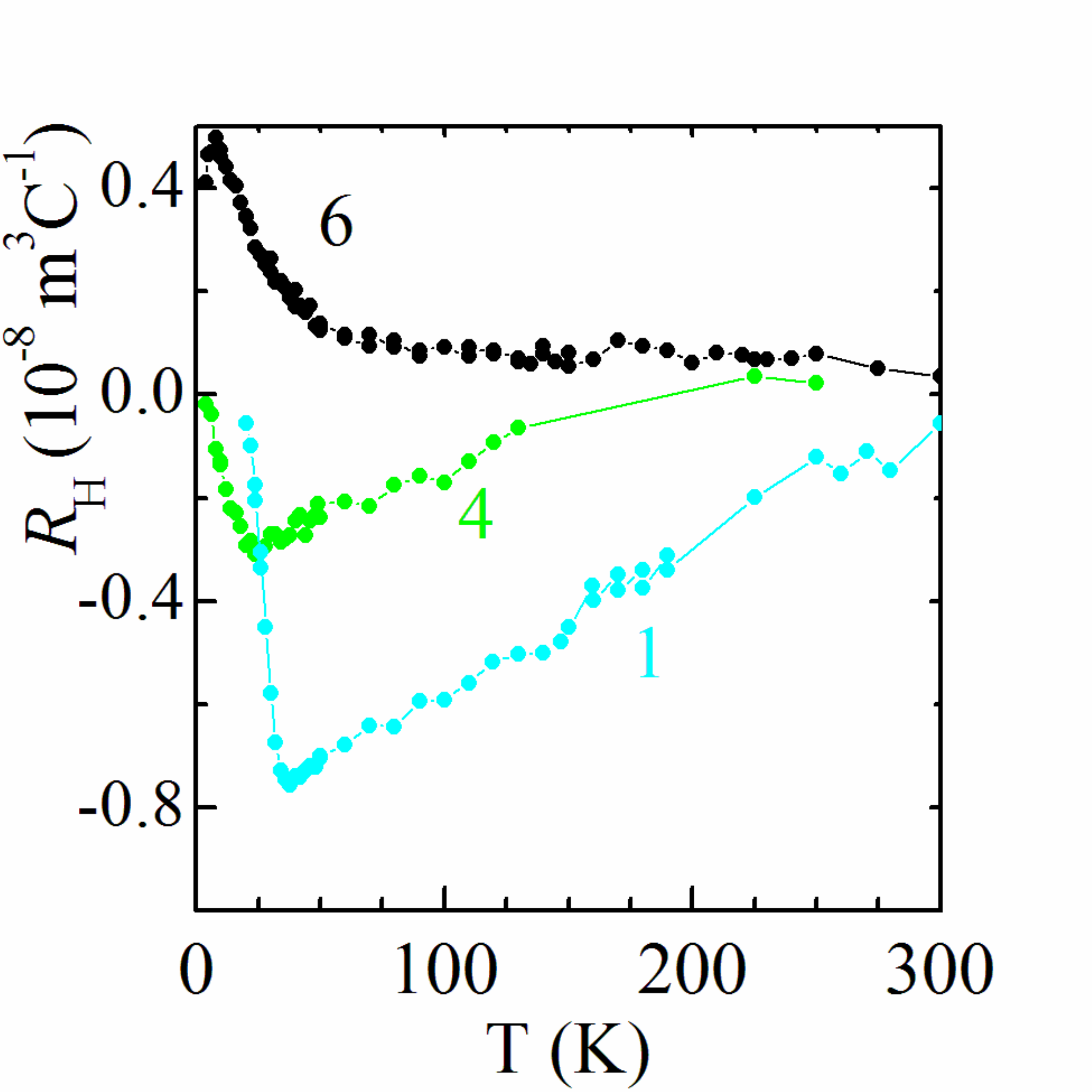}
\caption{Temperature dependence of Hall coefficient measured on three samples of Sr$_4$V$_2$O$_6$Fe$_2$As$_2$: Samples 1 and 4 are bulk superconductors, and Sample 6 is not.}
\label{fig:Hall}
\end{figure}
\begin{figure}[t]
\centering
\includegraphics[width=0.75\columnwidth]{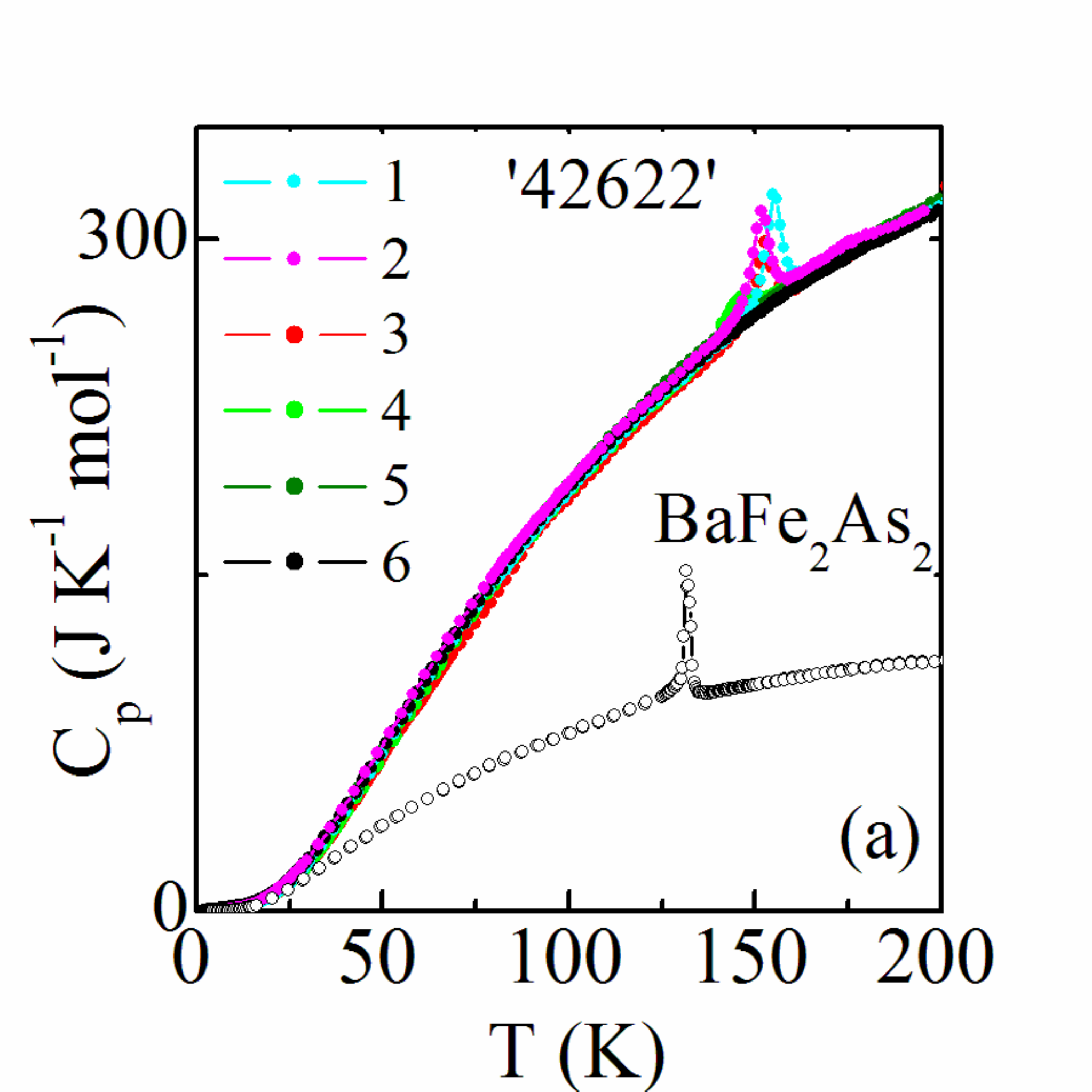}
\includegraphics[width=0.75\columnwidth]{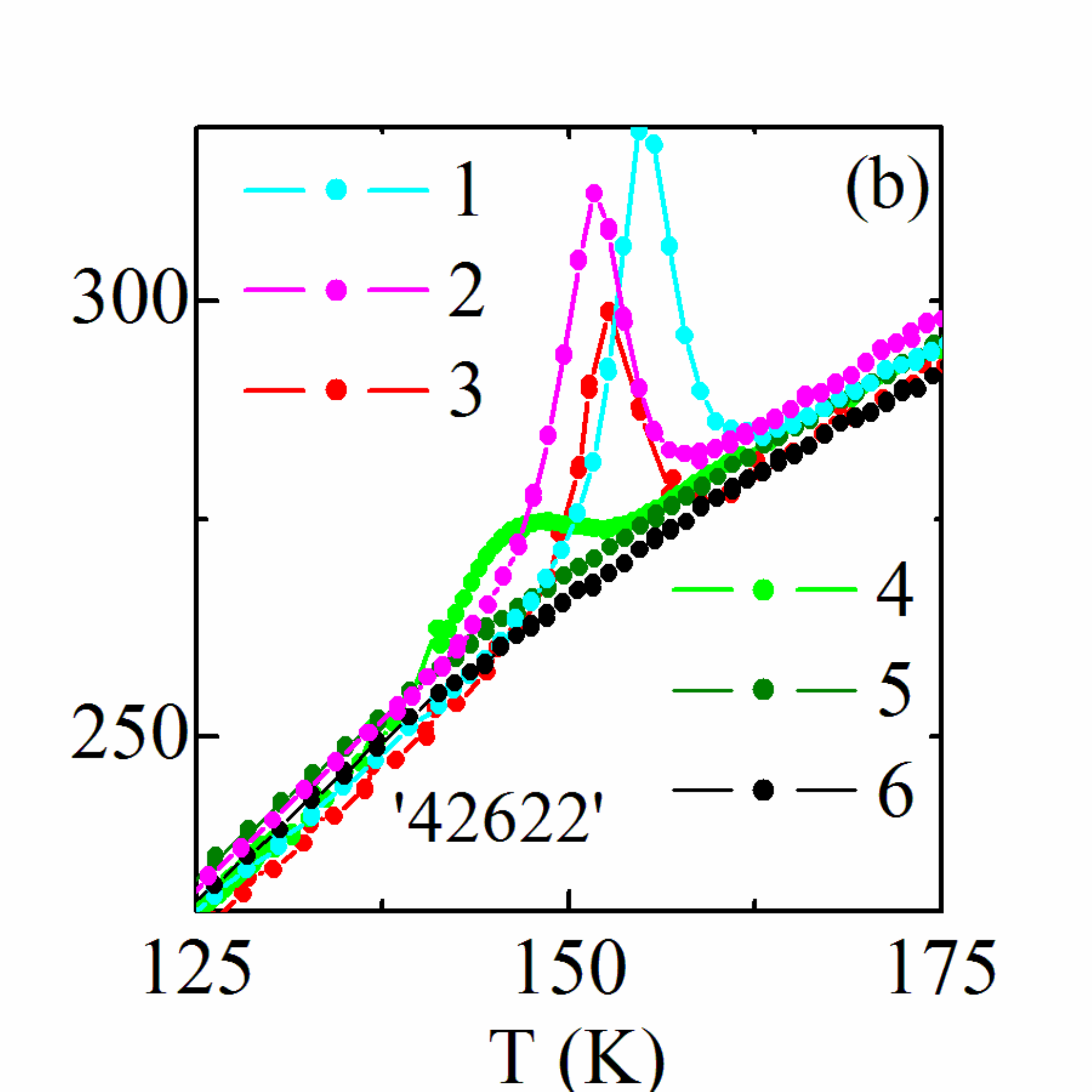}

\caption{Temperature variation of heat capacity ($C_p$) for various Sr$_4$V$_2$O$_6$Fe$_2$As$_2$ (42622) samples. The data for BaFe$_2$As$_2$ are shown in (a) for comparison with the 42622s. The enlarged $C_p$ region of 125 to 175 K for 42622s are shown in (b).}
\label{fig:Cp}
\end{figure}
The temperature-dependent Hall coefficients for Samples 1, 4 and 6 are shown in Figure~\ref{fig:Hall}. The conduction in these samples is either dominated by electron- or hole-like charge carriers. For each of the bulk superconductors, Samples 1 or 4, $R_H(T)$ is mainly negative with a sharp minimum at 38 K or 25 K, respectively. These results are consistent with that reported for
superconducting Sr$_4$V$_2$O$_6$Fe$_2$As$_2$ with $R_H = -1 \times 10^{-8}$ to $-0.7 \times 10^{-8}$ m$^3$/C for 50 to 160 K region.\cite{Zhu} For the non-superconductor (Sample 6), $R_H$ is positive with a sharp maximum near 7 K. It is interesting that $R_H$ changes sign, depending on the samples. 
\begin{figure}[t]
\centering
\includegraphics[width=0.75\columnwidth]{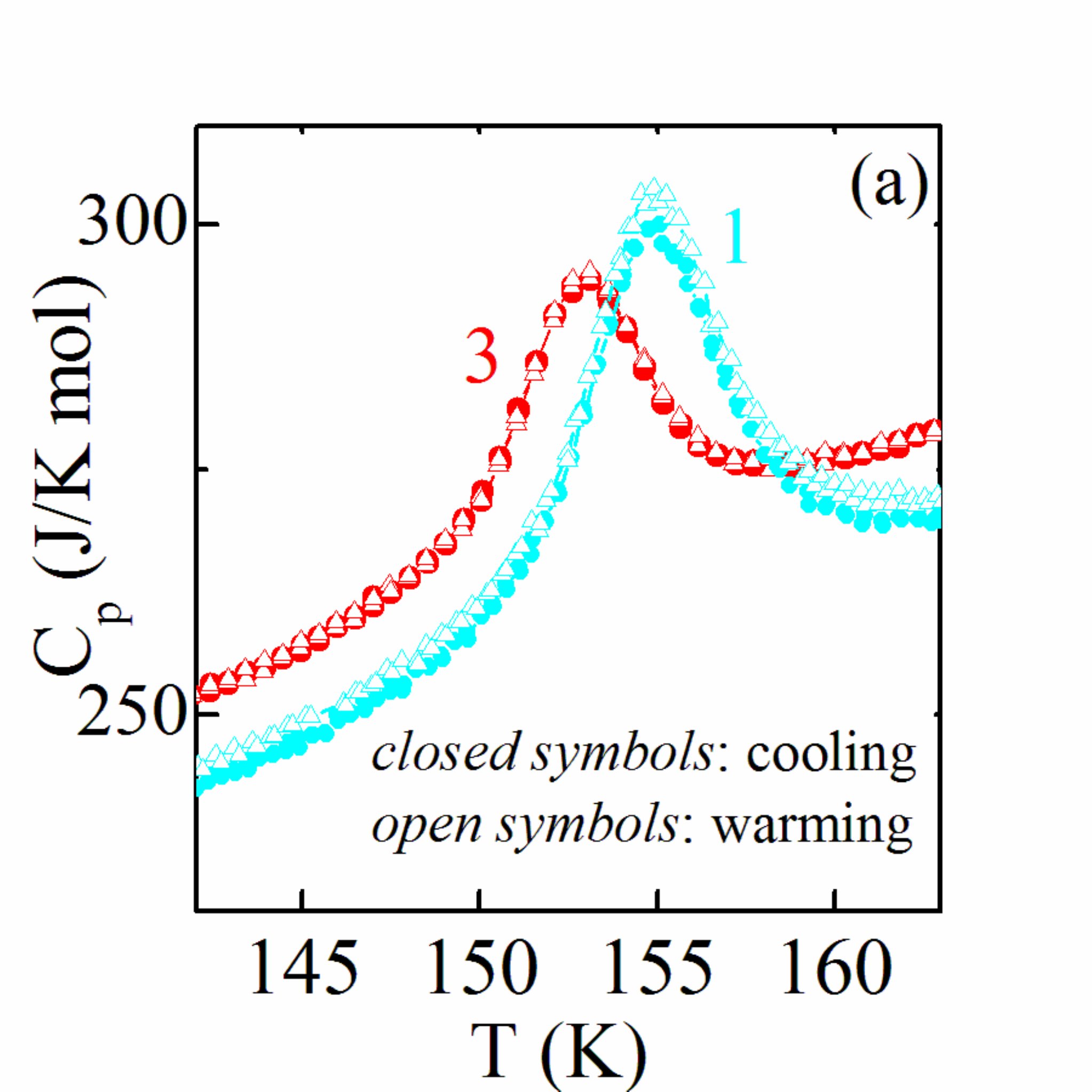}
\includegraphics[width=0.75\columnwidth]{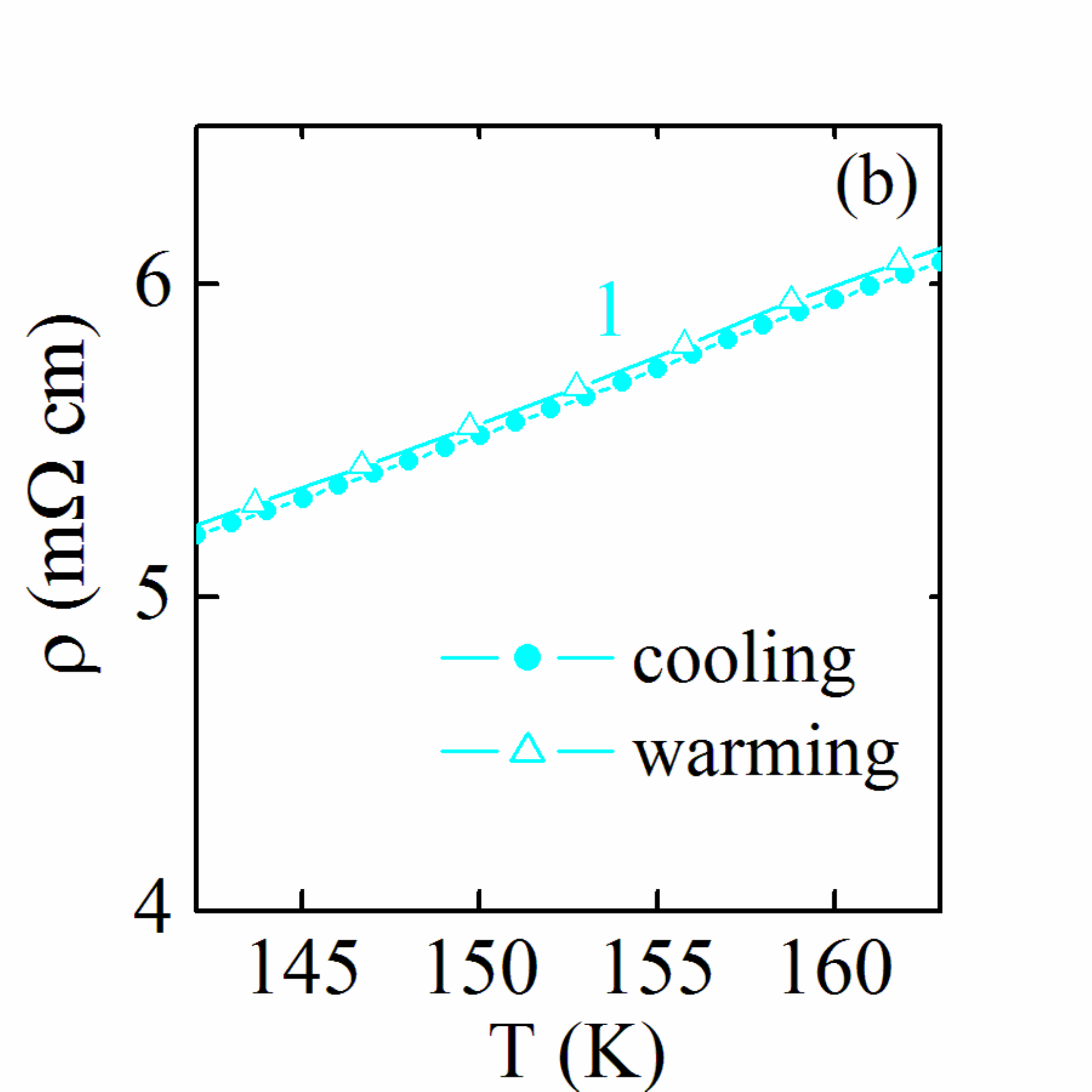}
\caption{(a) Temperature-dependent heat capacity measured upon cooling and then warming for
Samples 1 and 3. (b) Temperature-dependent electrical resistivity measured upon cooling and then
warming for Sample 1.}
\label{fig:coolwarm}
\end{figure}
In each, the magnitude of $R_H$ changes significantly across the measured temperature region (Fig.~\ref{fig:R}b). The temperature dependence of $R_H$ is probably caused by a multiband effect. In a multiband model, $R_H$ is the weighted sum of the contributions from each band and if the scattering rate of each band has different temperature dependence, then the weighted sum changes with temperature. The Fermi surfaces of iron-based superconductors consist of compensating holes and electron sheets. As such, the mobilities of different bands can have different temperature dependencies. Also, thermal expansion or slight difference in composition can produce a redistribution of the carriers among the bands so as to create a strong temperature dependence of $R_H$, as well as a change in sign of $R_H$. The inferred carrier concentrations at 100 K are $1.1 \times 10^{21}$ cm$^{-3}$ and $3.7 \times 10^{21}$ cm$^{-3}$ for Samples 1 and 4, respectively, and $6.8 \times 10^{21}$ cm$^{-3}$ for Sample 6, with the assumption of one-band model.

\begin{figure}[t]
\centering
\includegraphics[width=0.75\columnwidth]{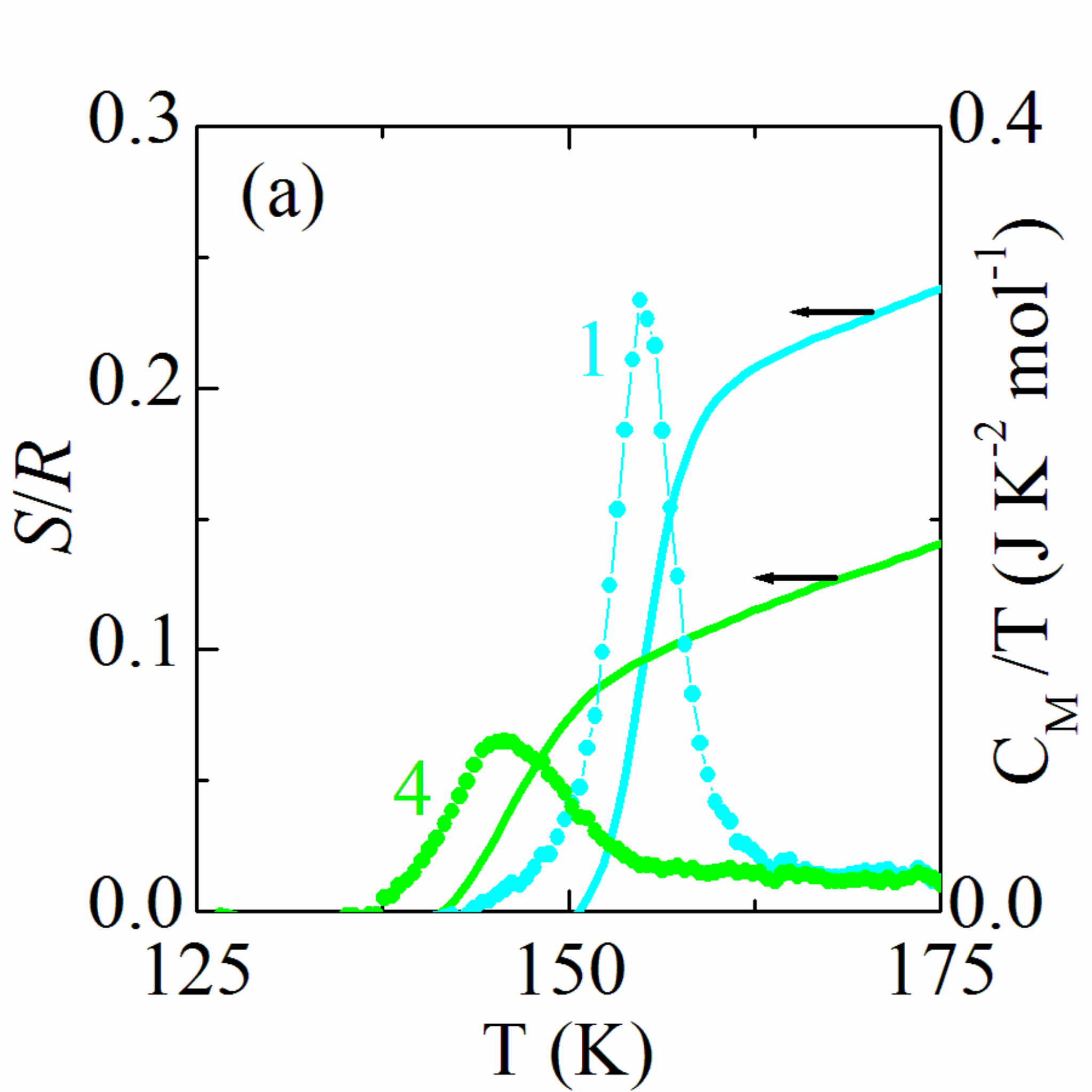}
\includegraphics[width=0.75\columnwidth]{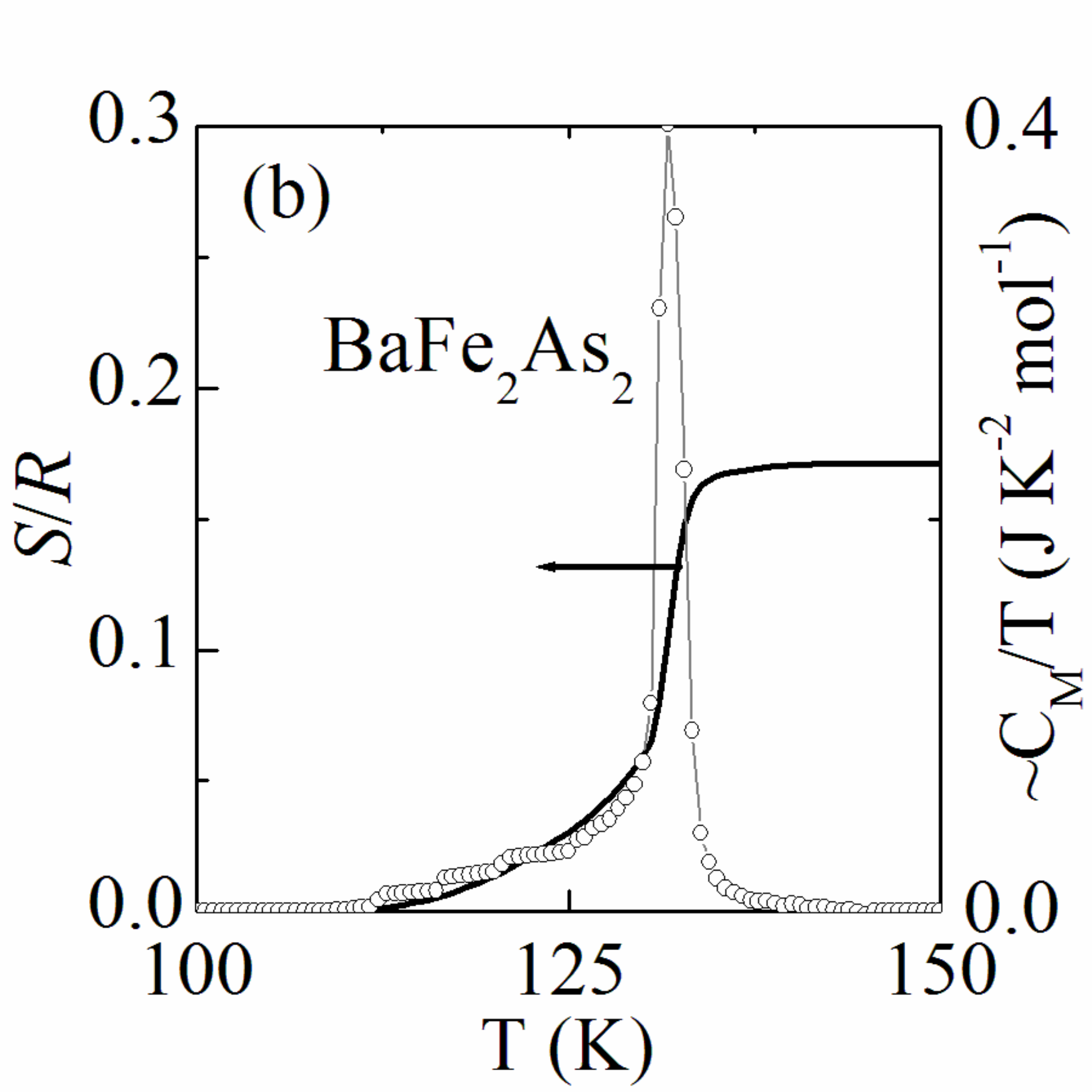}\\
\caption{Temperature dependence of magnetic specific heat in the form of $C_M/T$ and entropy change
($S/R$) for (a) Sr$_4$V$_2$O$_6$Fe$_2$As$_2$ Samples 1 and 4, and (b) BaFe$_2$As$_2$.}
\label{fig:entropy}
\end{figure}

The temperature-dependent heat capacity, $C_p(T)$, was measured upon cooling from 200 K to 1.8
K; the results are shown in Figure~\ref{fig:Cp}. There are only broad and barely visible features at $T_C$. However, sharp and large anomalies are seen at $\sim 150$ K for Samples 1, 2, and 3, while Sample 4 gives a broad feature. Non-bulk superconductors (Samples 5 and 6) show no peaks. Figure~\ref{fig:Cp}(a) also plots the heat capacity data for BaFe$_2$As$_2$ to show the relative magnitude of the high temperature peaks compared to 122's structural/spin-density-wave feature at $\sim 140$ K. Cao and
coworkers  found a roughly 150 K transition in the heat capacity in their ``overdoped'' 42622, and
declared it as an antiferromagnetic transition, that was followed by a weak ferromagnetic
transition at 55 K, and with a $T_C = 24$ K.\cite{Cao} Here we find that the 150 K transition is sample dependent, correlated with the superconductivity and that no transition is detected at $\sim 55$ K.

Such sharp transitions in heat capacity are likely to be intrinsic, although we suspected
impurities. Searching for likely binary and ternary impurity phases, only V$_2$O$_3$ is found to give a fairly sharp antiferromagnetic/structural transition in $C_p$ at $\sim 150$ K (upon cooling) that can be displaced to $\sim 175$ K (upon warming).~\cite{Keer} We checked for such possible thermal hysteresis in Samples 2 and 4, but did not find them (see Fig.~\ref{fig:coolwarm}a). $C_p$ was first measured upon cooling from 200 K down to 120 K, then upon warming for the same region. There is no peak displacement in heat capacity for these samples, suggesting second-order transitions. Temperature-dependent resistivity was measured on cooling from room temperature to 2 K, then on warming in the same region; this shows no feature (see Fig.~\ref{fig:coolwarm}b). Also, we have no evidence of a crystalline V$_2$O$_3$ impurity phase in x-ray and neutron diffraction results. Considering that such phase could exist amorphously, it would have to be the makeup of more than 5\% of each sample, in order to give the sharp feature in $C_p$; this is unlikely. Therefore, we assume that these high-temperature peaks are an intrinsic feature of the bulk superconducting 42622 samples.

In order to estimate the `magnetic' heat capacity ($C_M$) for Samples 1 and 4, the non-magnetic
contributions were approximated by Sample 6 (see Fig.~\ref{fig:Cp}b) and subtracted. The magnetic
entropies ($S$) were estimated by integration of $C_M/T$ versus $T$ and are shown in Figure~\ref{fig:entropy}(a). The magnetic entropy associated with each peak is small (comparable to BaFe$_2$As$_2$) and not recovered for the spin of 1/2 ground state for $\ln 2 = 0.69$ (Fig.~\ref{fig:entropy}b). This small entropy is expected to be associated with a small moment not detectable by neutrons ($\sim 0.1\mu_B$/vanadium).

\begin{figure}[t]
\centering
\includegraphics[width=0.75\columnwidth]{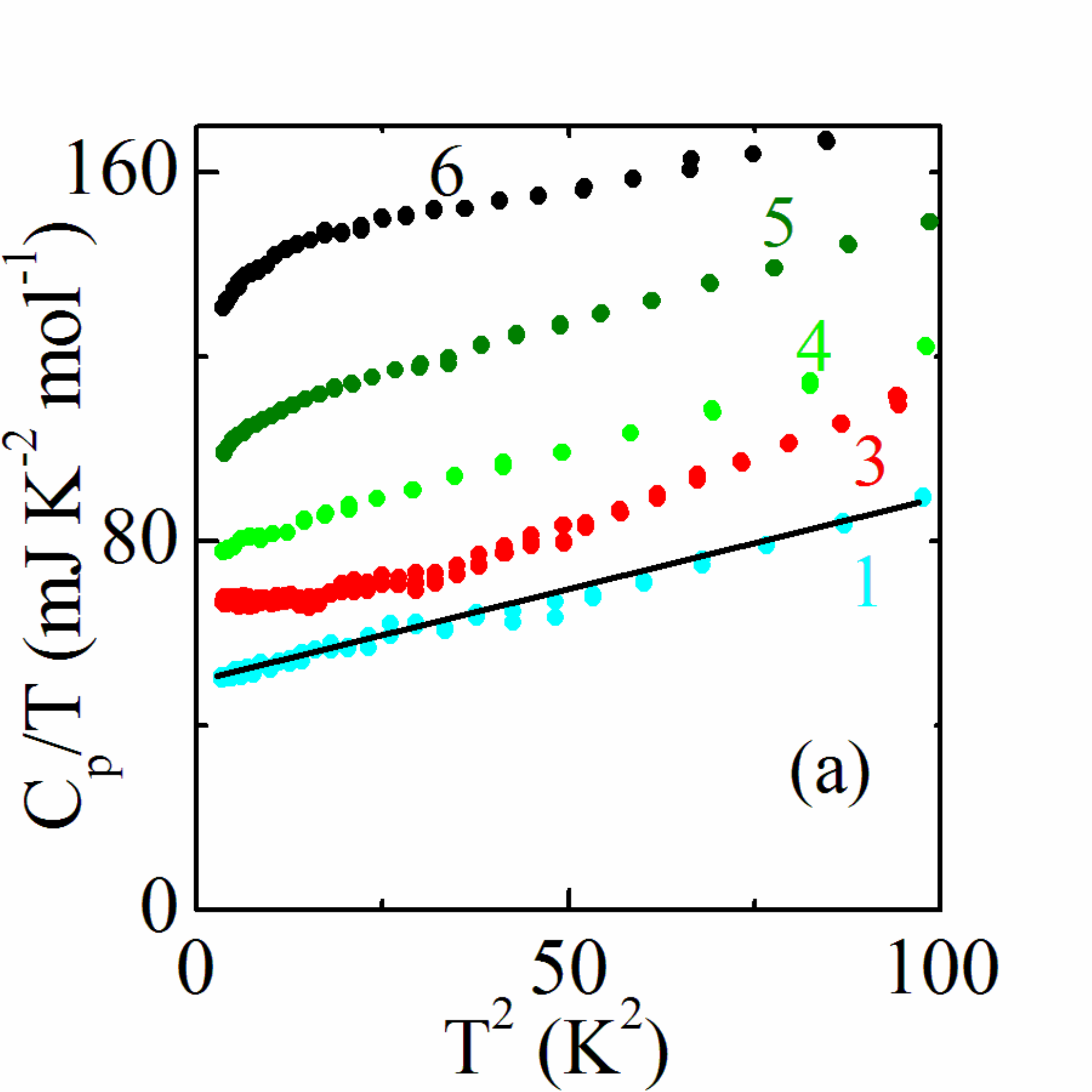}
\includegraphics[width=0.75\columnwidth]{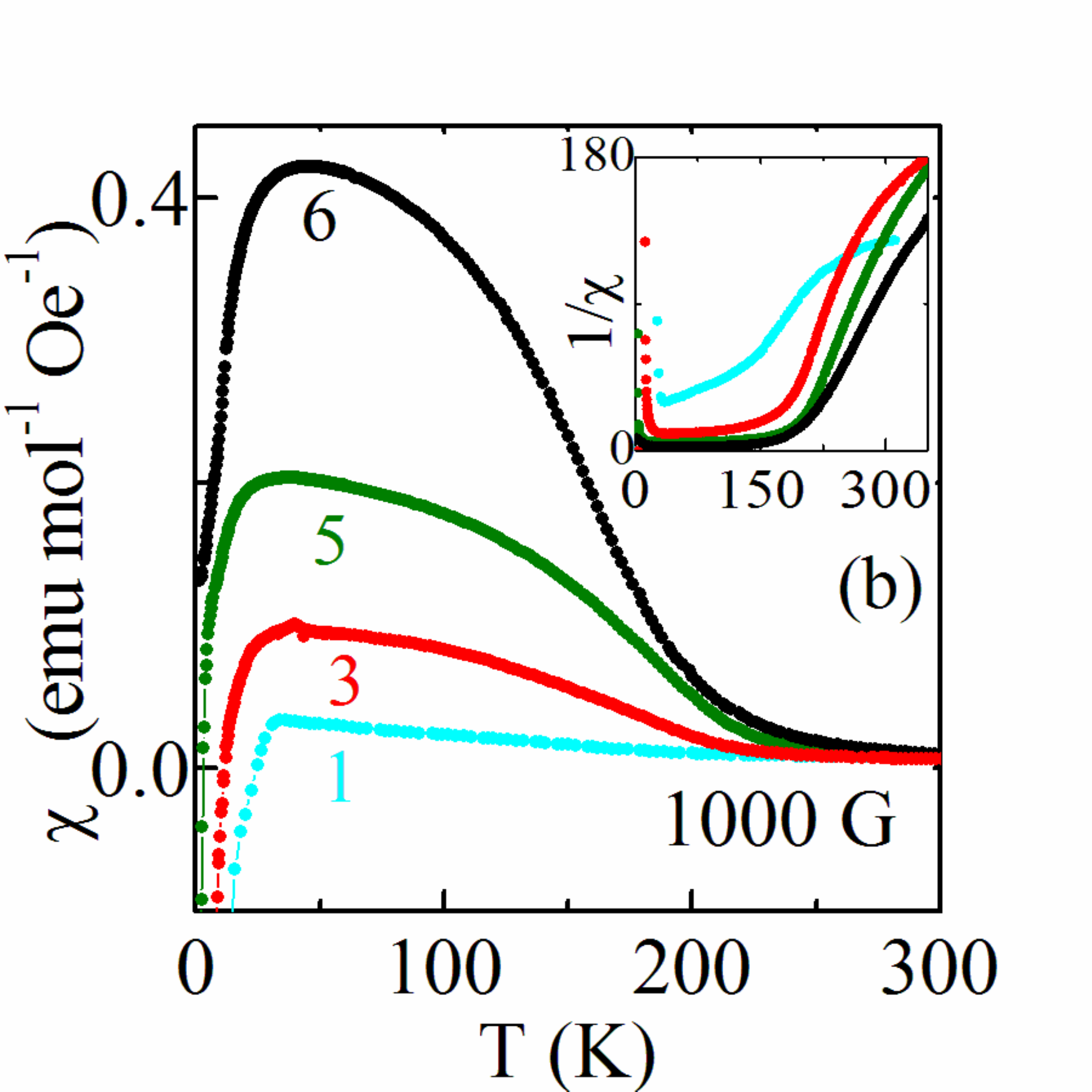}
\caption{For Sr$_4$V$_2$O$_6$Fe$_2$As$_2$ samples, temperature dependence of (a) specific heat is shown in $C_p/T$ versus $T^2$ form below 10 K, and also (b) magnetic susceptibility in 1000 G. The inset of (b) shows the non-linear behavior of inverse susceptibility.}
\label{fig:CpT2}
\end{figure}

Figure~\ref{fig:CpT2}(a) shows the temperature dependence of heat capacity for `Sr$_4$V$_2$O$_6$Fe$_2$As$_2$' Samples in the form of $C_p/T$ versus $T^2$ from 1.8 to 10 K. For Sample 1, the linear fit of this region yields a residual Sommerfeld coefficient of $\gamma_0 = 49.2$ mJ K$^{-2}$ mol$^{-1}$ (3.1 mJ K$^{-2}$ mol atom$^{-1}$). Although this region is less linear for the rest of samples, the $\gamma_0$ value may be estimated and it consistently goes
up with the Sample number (decrease in $T_C$). The $\gamma_0$ value increases with sample purity (see Fig.~\ref{fig:xray}), and is largest for Sample 6 at $\sim 125$ mJ K$^{-2}$ mol$^{-1}$. A close examination of magnetic susceptibility
results, displayed in Figure~\ref{fig:CpT2}(b) and~\ref{fig:suscept}(b), suggest the onset of a ferromagnetic-like ordering. Such spin correlations are consistent with the enhanced $\gamma_0$ and the destruction of superconductivity. It is noteworthy that our neutron diffraction measurements on Sample 6 would not be able to detect a very small ferromagnetic moment, due to the uncertainty in separating the magnetic from the
nuclear scattering. However, the presence of an intrinsic ferromagnetic order in our samples can
in principle be probed using polarized neutrons by applying a spin analysis of the scattered
neutrons; this will be the subject of a future study. These $\gamma_0$ values are much larger than those found in other Fe-based superconductors, for example LaFeAsO$_{0.89}$F$_{0.11}$ ($\gamma_0 = 4.1$ mJ K$^{-2}$ mol$^{-1}$) and BaFe$_{1.84}$Co$_{0.16}$As$_2$ ($\gamma_0 = 3.7$ mJ K$^{-2}$ mol$^{-1}$), or even the parent BaFe$_2$As$_2$ ($\gamma_0 = 6.1$ mJ K$^{-2}$ mol$^{-
1}$).

Our first principles calculations are similar to those reported by others.\cite{kwlee, mazin} In the LSDA and PBE calculations, the vanadium layers contribute states at the Fermi energy and the ground state is ferromagnetic at least within the V-O bilayers making up the perovskite part of the unit cell. It is possible that the ground state could be antiferromagnetic based on the $c$-axis stacking of ferrromagnetic bi-layers between different unit cells or that there could be a spiral or more complicated antiferromagnetic state. We note that because of the metallic nature of FeAs planes, long-range magnetic interactions (e.g. RKKY type) should be present. Also, as noted by Mazin
,\cite{mazin}  the material is more two dimensional than other Fe-based superconductors, and the energy differences between different magnetic states are small, both of which may depress and even
entirely suppress the ordering temperature. In any case, the ground state for Sr$_4$V$_2$O$_6$Fe$_2$As$_2$ is either ferromagnetic or close to ferromagnetism at this level of theory.

Of course, ferromagnetism is generally incompatible with singlet superconductivity, and even
ferromagnetic fluctuations are expected to be strongly pair breaking. Therefore even though, as
discussed by Mazin,~\cite{mazin} the Fermi surface as related to the FeAs layers retains the features
thought to be essential for superconductivity, the presence of ferromagnetic or near
ferromagnetic metallic V-O layers is expected to be highly antagonistic to superconductivity
(note that scattering between the two subsystems, even if they are quite distinct would be pair
breaking). In the PBE+U calculations, the V $d$-states are gapped away from the Fermi energy
leaving the Fe bands and the ground state as antiferromagnetic. For the non-spin-polarized state
with the PBE functional, we obtain a total electronic density of states, $N(E_F)=18.1$ eV$^{-1}$ per unit cell, mostly from the vanadium. This high value strongly favors magnetism. It corresponds to a bare specific heat coefficient, $\gamma_b = 43$ mJ K$^{-2}$ mol$^{-1}$ (mole of two V atoms unit cell). With ferromagnetism we obtain $N(E_F) = 9.6$ eV$^{-1}$ per cell, $\gamma_b = 22.7$ mJ K$^{-2}$ mol$^{-1}$. For the PBE+U calculations, the vanadium carries moments close to the expected $S = 1$ value, in particular $1.75\:\mu_B$ within the vanadium LAPW sphere of radius 2.05 Bohr. We find that $N(E_F)$ is essentially independent of the vanadium magnetic order, comes from the Fe and has a lower value of 2.3 eV$^{-1}$ per cell, corresponding to $\gamma_b = 10.7$ mJ K$^{-2}$ mol$^{-1}$. The electronic structure has two electrons in the majority spin $t_{2g}$ orbitals, which can accommodate three electrons. This orbital degeneracy is what leads to orbital ordering in V$^{3+}$ compounds. Such ordering is possible in this compound, but the details will depend on the exact crystal structure including atomic positions, which is not known at present. Comparing these values for the specific heat with the experimental data, we 
conclude that there are substantial enhancements above the bare values in both states.
Nonetheless we associate the high specific heat coefficients measured in non-superconducting
samples, with metallic vanadium perhaps strongly enhanced by correlations, and the lower
values in superconducting samples with non-metallic, insulating V.

\section{Conclusions}
The Sr$_4$V$_2$O$_6$Fe$_2$As$_2$ compound is much harder to synthesize than other families and single
crystals will be crucial for determining intrinsic behavior. We have found that the $T_C$ value is
high for samples with most crystalline impurities, and it decreases with increases in $c$-lattice
parameter. There are no structural transitions in these compounds, but there is some evidence of
magnetic order. We believe that we can tune the V-O subsystem from a correlated metallic state
to an insulating state, and that there is an interplay between the electronic behavior of the V-O
subsystem and the FeAs superconductivity. This interplay is presumably magnetic in nature. It
will be of considerable interest to elucidate the character of magnetism associated with
vanadium, both in terms of ordering and spin fluctuations, especially via polarized neutron
scattering measurements. The results suggest that Sr$_4$V$_2$O$_6$Fe$_2$As$_2$ is a compound that offers a window into the interplay of Fe-based superconductivity and correlated oxide physics with
tunability through chemical stoichiometry.

\section{Acknowledgments}

Research at ORNL is sponsored by the Materials Sciences and Engineering Division, Office of
Basic Energy Sciences, US Department of Energy. We acknowledge discussions with D. K.
Christen, V. Keppens and D. Mandrus for technical support. The work at High Flux Reactor
(ORNL) was sponsored by the Scientific User Facilities Division, Office of Basic Sciences, U. S.
DOE.

\end{document}